\documentclass[aps,showpacs,twocolumn,amsmath,longbibliography]{revtex4-1}
\usepackage{amsfonts}
\usepackage{hyperref}
\usepackage{graphicx}
\usepackage{float}
\usepackage{xcolor}
\usepackage[normalem]{ulem}
\usepackage[utf8]{inputenc}

\begin{document}

\title{Non-orthogonal eigenvectors, fluctuation-dissipation relations and entropy production}
\author{Yan V. Fyodorov}
\affiliation{King's College London,  Department of Mathematics, London  WC2R 2LS, United Kingdom}
\author{Ewa Gudowska-Nowak}
\affiliation{Institute of Theoretical Physics and Mark Kac Center for Complex Systems Research, Jagiellonian University, Krak\'ow, Poland}
\author{Maciej A. Nowak}
\affiliation{Institute of Theoretical Physics and Mark Kac Center for Complex Systems Research, Jagiellonian University in Krak\'ow, Poland}
\author{Wojciech Tarnowski}
\affiliation{Institute of Theoretical Physics and Mark Kac Center for Complex Systems Research,  Jagiellonian University in Krak\'ow, Poland}
\begin{abstract}

 Celebrated fluctuation-dissipation theorem (FDT) 
linking the response function to time dependent correlations of observables measured in the reference unperturbed state is one of the central results in equilibrium statistical mechanics. 
In this letter we discuss an extension of  the standard FDT to the case when multidimensional matrix representing transition probabilities is strictly non-normal. This feature dramatically modifies  the dynamics, by incorporating the effect of  eigenvector non-orthogonality via the associated overlap matrix of Chalker-Mehlig type. In particular, the rate of entropy production per unit time is strongly enhanced by that matrix.   We suggest, that this mechanism  has an impact on the studies of collective phenomena in neural matrix models, leading, via transient behavior, to such phenomena as synchronisation and emergence of the memory. We also expect,  that the described mechanism generating the entropy  production is generic for wide class of  phenomena, where dynamics is driven by non-normal operators. For the case of driving by a large  Ginibre matrix the entropy production rate is evaluated analytically, as well as for the Rajan-Abbott model for neural networks.

\end{abstract}

\newcommand{\dg}{\dagger}
\newcommand{\tr}{\textup{Tr}}
\newcommand{\btr}{\textup{bTr}}
\newcommand{\<}{\left<}
\renewcommand{\>}{\right>}
\newcommand{\idm}{\mathbf{1}}
\newcommand{\zb}{\bar{z}}
\newcommand{\wb}{\bar{w}}
\newcommand{\vb}{\bar{v}}
\newcommand{\gb}{\bar{g}}
\newcommand{\linia}{\rule{\linewidth}{0.4mm}}
\newcommand{\unit}{\mathbb{1}}
\newcommand{\cG}{{\cal G}}
\newcommand{\cQ}{{\cal Q}}
\newcommand{\cX}{{\cal X}}
\newcommand{\cY}{{\cal Y}}
\newcommand{\cA}{{\cal A}}
\newcommand{\cR}{{\cal R}}
\newcommand{\cB}{{\cal B}}
\newcommand{\cW}{{\cal W}}
\newcommand{\cL}{{\cal L}}
\newcommand{\cD}{{\cal D}}
\newcommand{\ket}[1]{\left| #1 \>}
\newcommand{\bra}[1]{\< #1 \right|}
\newcommand{\braket}[2]{\< #1 | #2 \>}
\newcommand{\bb}[1]{{\bf #1}}
\newcommand{\re}{\textup{Re}}
\newcommand{\im}{\textup{Im}}
\newcommand{\be}{\begin{eqnarray}}
\newcommand{\ee}{\end{eqnarray}}

\newcommand{\MeijerG}[8][\bigg]{G^{{ #2 },{ #3 }}_{{ #4 },{ #5 }} #1( \begin{matrix} #6 \\ #7 \end{matrix}\, #1\vert\, #8 #1)}

\maketitle

Close to equilibrium the regression hypothesis and linear response theory allow computation of transport coefficients (e.g. diffusion, mobility, electric conductivity) in terms of unperturbed averages of total current correlation functions \cite{Zwanzig,Livi,Kubo12,Vulp2008,Callen,Reggiani20}. This relationship between (in general, time-dependent) forcing and time correlation of observables of interest in perturbed systems is a fundamental result of statistical physics known as the fluctuation-dissipation theorem (FDT)~\cite{Kubo_1966}. The theorem states that if a system fluctuates in the presence of a restoring linear dissipative force, the variance of fluctuations - which is a stationary property - is linearly related to the dissipation coefficient which features the transient response of the system.
Developments of response theory to far-from-equilibrium phenomena - like slow relaxing structural glasses and proteins, mesoscopic radiative heat transfer or driven granular media where the violation of FDT was observed  \cite{Ricci-Tersenghi00,Crisanti03,Barrat98,Bellon02,Hayashi07,Perez-Madrid09,Averin10} initiated interest in various generalizations and interpretations of FDT to systems with non-zero entropy production. In particular, for growth phenomena such as those described by the Kardar-Parisi-Zhang equation~\cite{Kardar86} different formulations of the FDT have been proposed~\cite{Wio17,Rodriguez19}. New approaches have been brought forward  to address fluctuation-dissipation relations in systems exhibiting anomalous transport properties, including sub-and super-diffusion~\cite{Pottier2004,Ewa2,Dybiec12} and derivation of response relations from a trajectory-based description has been advanced~\cite{Maes20,Maes2020} along with investigation of "dynamical activity" as a complement to entropy,
  with the excess in the latter quantity propounded as a new Lyapunov functional.
 In recent years functionality and use of FDT and its generalized forms has been assessed in various fields: information theory~\cite{Hauke}, quantum chaos~\cite{Tsuji2018, Pappalardi2022},  climatology \cite{Hassa,Nicolis}, inviscid hydrodynamics, active matter \cite{Prost,Sar2021}
and analysis of spontaneous and stimulus-related activity of neuronal networks \cite{VanMeegen,neural_nemen}, experimental retrieval of Green's functions in complex reverberating enclosures \cite{Davy_2013,Davy_2016}, to name just a few.\\

\noindent The aim of this work is to shed new light on a generalization of fluctuation-dissipation relation for a multivariate dynamical system whose linearized dynamics close to a stationary state is described by the set of Langevin equations 
\begin{equation}
\dot{x}_i(t)=\sum_{j=1}^{N}A_{ij}x_j(t)+\xi_j(t),
\label{eq:lang}
\end{equation}
where the driving matrix $A$ is large, random  and not symmetric and, in particular, not normal. Matrix $A$ is called normal if $AA^T = A^TA$. Here $\xi(t)$ represents a random (Gaussian) driving force  with zero mean and the correlation matrix $\<\xi_i(t)\xi_j(t')\>=B_{ij}\delta(t-t')$.
This process can be also viewed as a multivariate version of the Ornstein-Uhlenbeck process~\cite{WangUhlenbeck,Zwanzig,Godreche_2019}.
The multivariate Langevin dynamics is relevant in many instances, e.g. in modeling Markovian reaction networks, with entries of the matrix $A$ representing transition rates between states derived from the chemical master equation~\cite{Chen_2022} or dynamic response and localization phenomena in population dynamics \cite{May} and biological transport \cite{Sar2021}. For neural networks matrix $A$ captures synaptic couplings between neurons~\cite{Sompolinsky_1988,Enzo_2023}. Not dissimilar equations can  also be shown \cite{Fyod_unpub} to underlie the description of diffusive fields in reverberating cavities in the framework of effective Hamiltonian approach~\cite{Davy_2021}.\\
We start our consideration with recalling  a formulation of FDT in the form known as a continuous Lyapunov (Sylvester) equation. Then, we discuss modifications necessary when the asymmetric coupling matrix $A$,  despite having a complex spectrum, can be diagonalized by a single unitary transformation. Such a situation of {\it normal} matrices is however non-generic, which motivates us to consider a generic case of truly non-normal  $A$. We then   present the generalized FDT, pointing out the crucial  role played for the  system's dynamics by the left-right eigenvector non-orthogonality factors introduced by  Chalker and Mehlig~\cite{ChalkerMehlig}. Next, we link those factors with the entropy production rate~\cite{Godreche_2019}, and show that increasing non-normality in the system, while keeping spectrum unaltered enhances entropy production.  Our numerical results confirm the importance of eigenvector non-orthogonality  and display the associated transient behavior. We conclude by discussing  the relevance of the proposed generalization not only for the study of real neuronal systems, but also for much broader class of models with non-normality built in.  \\
We assume that $A$ is a time-independent matrix whose all eigenvalues have only  negative real parts to ensure stability of the steady state $x=0$.
The general solution of (\ref{eq:lang}) with the initial condition $x(0)=x_0$ is $x(t)=e^{At}x_0+\int_0^t ds e^{A(t-s)}\xi(s)$ and the covariance of instantaneous fluctuations
 $\delta x(t)=x(t)-e^{At}x_0$ is given by the $N\times N$ matrix
\begin{eqnarray}
C(\tau,t)=\<\delta x(t+\tau)\delta x^T (t)\>=e^{A\tau}C(0,t), 
\end{eqnarray}
where we assumed $\tau>0$. The equal-time covariance matrix given by
\begin{eqnarray}
 C(0,t)= \int_0^t e^{As}Be^{A^Ts}ds.
\end{eqnarray}
In the large time limit it reduces to $C_0\equiv C(0,\infty)$ that satisfies the Lyapunov  equation~\cite{Zwanzig,Fox1979,Onsager31,Prost,Landi_2013,Sar2019}
\begin{equation}
AC_0+C_0A^T=-B \label{eq:CO}
\end{equation}
and the correlation function defined for arbitrary $\tau\in \mathbb{R}$  reads
\begin{equation}
C(\tau)=e^{A\tau}C_0\theta(\tau)+C_0e^{-A^T\tau}\theta(-\tau),\label{eq:CFull}
\end{equation}
which implies $C(-\tau)=C^T(\tau)$.\\
Eq.~\eqref{eq:CO} expresses the celebrated FDT for a linear dissipative dynamics balancing 
restoring action of the coupling matrix $A$ with strength of random driving encoded in $B$. In Supplemental Material~\cite{SUP}, we show how to relate  Eq.(4) to the classical Einstein's formulation of FDT.\\
We discuss now the generic case when matrix $A$ is non-normal, i.e. does not commute with its transpose. 
Then, in general, (\ref{eq:CO}) can be solved 
by using the spectral decomposition $A=\sum_{k=1}^{N}\ket{R_k}\lambda_k\bra{L_k}$ in terms of 
 right (left)  eigenvectors $\ket{R_i}$ and $ \bra{L_i}$   corresponding to the {\it  same} complex eigenvalue $\lambda_i$. For the transposed matrix this implies $A^T=\sum_{k=1}^{N}\overline{\ket{L_k}}\lambda_k\overline{\bra{R_k}}=\sum_{k=1}^{N}\ket{L_k}\bar{\lambda}_k\bra{R_k}$, which allows one to convert  \eqref{eq:CO} into
\begin{equation}
C_0=\int_0^{\infty}e^{As}B e^{A^Ts}ds=-\sum_{k,l=1}^{N}\frac{\ket{R_k}\bra{L_k}B\ket{L_l}\bra{R_l}}{\lambda_k+\bar{\lambda}_l}. \label{eq:C0SpecDec}
\end{equation}
Differentiating \eqref{eq:CFull} and making use of \eqref{eq:CO} leads to FDT expressed as
\begin{equation}
2\partial_{\tau}C(\tau)=-\chi(\tau)B+B\chi^T(-\tau)+\Delta(\tau), \label{eq:FDTReal}
\end{equation} 
where $\chi(t)=e^{At}\theta(t)$ stands for the response function and the "FDT violation" term
\begin{equation}\label{FDTviol}
\Delta(\tau)=AC(\tau)-C(\tau)A^T
\end{equation}
is induced by the asymmetry of $A$, as we shall discuss in details below.  
The above analysis can be also performed in Fourier space (Supplemental Material~\cite{SUP}).\\
To further quantify the effects of non-normality, we rewrite $\Delta(\tau)$ using the spectral decomposition of $A$: 
\begin{eqnarray}
\Delta(\tau)=-\sum_{k,l}\ket{R_k}\bra{L_k}B\ket{L_l}\bra{R_l}\nonumber\\ \times \frac{\lambda_k-\bar{\lambda}_l}{\lambda_k+\bar{\lambda}_l}\left(e^{\lambda_k\tau}\theta(\tau)+e^{-\bar{\lambda}_l\tau}\theta(-\tau)\right).
\end{eqnarray}
We remark here that a similar observation  was made in~\cite{Godreche_2019}. 
In case of real symmetric $A$  and $B=\idm$ the eigenvectors are orthogonal and eigenvalues are real, which with $\bra{L_k}B\ket{L_l}=\delta_{kl}$ implies $\Delta(\tau)=0$.
For asymmetric but normal $A$ with $B=\idm$ the eigenvalues are complex, but eigenvectors are still orthogonal so that
\begin{equation}
\Delta(\tau)=-\sum_{k=1}^{N}\ket{R_k}\bra{R_k}\frac{i \im\lambda_k}{\re\lambda_k}\left(e^{\lambda_k\tau}\theta(\tau)+e^{-\bar{\lambda}_k\tau}\theta(-\tau)\right),
\label{eq:DeltaNormal}
\end{equation}
which means that the different eigenmodes of $A$ contribute independently to $\Delta(\tau)$.\\
We present now the main result of this paper. 
For a generic non-normal $A$ and uncorrelated  noise $B=\idm$ it is convenient to introduce the traced object $\tilde{\Delta}=\frac{1}{N}\tr\Delta$, which simplifies the formulas thanks to $\tr\ket{R_k}\braket{L_k}{L_l}\bra{R_l}=\braket{L_k}{L_l}\braket{R_l}{R_k}=O_{kl}$. The matrix ${\cal O}$ with entries $O_{kl}$ encodes the left-right eigenvector overlaps and the difference 
${\cal O}-\idm$ is a convenient standard indicator of non-normality. In such a case we arrive at
\begin{equation}
\tilde{\Delta}(\tau)=-\frac{1}{N}\sum_{kl}O_{kl}\frac{\lambda_k-\bar{\lambda}_l}{\lambda_k+\bar{\lambda}_l}\left(e^{\lambda_k\tau}\theta(\tau)+e^{-\bar{\lambda}_l\tau}\theta(-\tau)\right)
\label{eq:DeltaTilde}
\end{equation}
making evident  that the contributions of different eigenmodes  are strongly coupled through the matrix of eigenvector overlaps.
The appearance of the overlap matrix ${\cal O}$ in the above equation is remarkable.
First of all,  such a matrix is known to control the spectral stability of linear systems
under a small additive perturbation $\delta A$ of the coupling matrix $A$. The leading change in matrix spectrum is controlled by the first-order perturbation  $\delta \lambda_i= \bra{L_i}\delta A\ket{R_i} \le \sqrt{\braket{L_i}{L_i} \braket{R_i}{R_i}} ||\delta A||_2$, where $||\delta A||_2$ denotes the Frobenius norm and the last inequality follows from Cauchy-Schwarz inequality.  The square root of the diagonal entries $O_{ii}\geq  1$ featuring in the above estimate is known in numerical analysis community as the eigenvalue condition number determining the enhanced sensitivity of the perturbation in the case of non-normal operators. It is instructive to  calculate $\tilde{\Delta}=\frac{1}{N}\tr\Delta$ with \eqref{eq:DeltaNormal} and compare the result  to \eqref{eq:DeltaTilde}.  
The matrix ${\cal O}$ also has  an impact on transient behavior  and other characteristics of systems with classical dynamics \cite{Grela_2017,erdos2018power,Tarnowski_2020,Gudowska_neuro}  like  relaxation, scattering and other properties of their quantum counterparts \cite{Savin,schomerus2000quantum,davy2019probing,fyodorov2022eigenfunction,cipolloni2023non,cipolloni2023entan,ghosh2023eigenvector}. All this has motivated a long-standing interest in statistics of entries $O_{ij}$ of the overlap matrix for random linear operators and matrices \cite{ChalkerMehlig,janik1999correlations,burda2014dysonian,fyodorovmehlig2002,fyodorovSavin2012,fyodorov2018CMP,bourgadedoubach2020,dubach2021,fyodorov2021condition,WFC1}.\\
Considering   the formal solution to noiseless~\eqref{eq:lang}  for the initial vector "spike" $x(t)=x(0) \delta(t)$ and  decomposing $A= X-\mu \idm $ to explicitly include May's stability condition, the spectral decomposition of $X$ in terms of its left and right eigenvectors yields the following squared norm of the solution $D(t)=\braket{x(t)}{x(t)}$ known as the Euclidean distance from the stationary point:
\begin{equation}
D(t) =e^{-2\mu t} \sum_{i,j}^N\braket{x(0)}{L_i}\braket{R_i}{R_j}\braket{L_j}{x(0)} e^{t(\bar{\lambda}_i +\lambda_j)} \label{dist} 
\end{equation}
Further assuming spikes to be  uniformly distributed over  $N-$ dimensional hypersphere  leads upon averaging to 
\begin{equation}
\tilde{D}(t)=e^{-2\mu t}   \sum_{k,l} e^{t (\lambda_k+ \bar{\lambda}_l)}O_{lk}
\end{equation}
We again see that different eigenvalues  are coupled  during the evolution, and the coupling is further strengthened  by eigenvectors overlap matrix. As we discuss below,  such coupling is known to generate a transient behavior for $\tilde{D}(t)$.\\
We can strengthen this argument by deriving the explicit formulae  for entropy production rate per unit time  $\Phi$ in terms of the eigenvector overlaps. 
Non-zero entropy production rate is equivalent to irreversibility, and play crucial role in every thermodynamic analysis of non-equilibrium processes. Since $\Phi$ can be expressed 
as \cite{Qian,Landi_2013,Godreche_2019}
\begin{eqnarray}
    \Phi\equiv 2\int \frac{{\bf{J}}^T({\bf{x}}){\bf{B}}^{-1}{\bf{J}}({\bf{x}})}{P({\bf{x}})}d{\bf{x}}=-{\rm Tr}\, B^{-1}A\Delta(0),
    \label{eq:EntropyProd1}
\end{eqnarray}
where $P({\bf{x}})$, ${\bf{J}}$ stand for the stationary probability density and stationary probability flux, respectively (in Supplemental Material~\cite{SUP}, we derive the form of  \eqref{eq:EntropyProd1} from Smoluchowski-Fokker-Planck equation),
the explicit calculation of the trace yields
\begin{eqnarray}
 \Phi=\sum_{k,l} \bra{R_l}B^{-1}\ket{R_k}\bra{L_k}B\ket{L_l} \lambda_k \frac{\lambda_k-\bar{\lambda}_l}{\lambda_k+\bar{\lambda}_l}
\end{eqnarray}
Note that for the class of linear transformations of initial multivariate variables $x_i$, similarity transformation allows to choose $B=\bf{1}$, 
and preserves complex eigenvalues of $A$. This observation allows us to write the  entropy production rate as 
\begin{eqnarray}\label{phi}
\Phi=\sum_{k,l} O_{kl} \Lambda_{kl} 
\end{eqnarray}
where $\Lambda_{kl}=\lambda_k(\lambda_k-\bar{\lambda}_l)/(\lambda_k+\bar{\lambda}_l)$. 
 This expression elucidates the  main role of non-orthogonality factors $O_{kl}$ in the rate of entropy production.  Not only those factors couple different eigenmodes, but also enhance the entropy production dramatically, as the diagonal overlaps strongly grow  with the  dimension of the multivariate vector $N$. When the value of index $N$ is small a brute force double summation in (\ref{phi}) is possible.  On the other hand, when $N$ is large and $X$ is random, another approach, relying on the properties of stochastic operators  in the leading planar approximation,  turns out to be very efficient. Particularly useful is the introduction  of large $N$ resolvents of the type $W_1(z,w)\equiv\frac{1}{N}\<{\rm Tr} \frac{1}{z-X} \frac{1}{w-X^T}\>$ and applications of Dunford-Taylor formulae,  as discussed in Supplemental Material~\cite{SUP}.  
 These techniques allow   calculation of entropy production rate in a \textit{non-normal} linear system driven by the  Ginibre ensemble, as well as in the Rajan-Abbott model for neural networks, discussed beneath. Such results seem not to be available in the existing literature.\\
To elucidate the implications of eigenvector non-orthogonality factors on a concrete example, we focus on the dynamical models of neuron networks. Understanding of dynamics of signal propagation through such networks and the emergence of collective long-time phenomena like synchronisation and/or memory still represents a formidable challenge. 
In what follows we  choose $A$ to be a random matrix and explore spectral properties of the  covariance matrix $C_0$ satisfying  \eqref{eq:CO}.  As a justification of our choice, we recall that random matrices  appear naturally as couplings in several models or real neural networks or artificial networks in machine learning~\cite{Pennington}. \\
In below, we juxtapose  three different cases of randomness: (i) the normal matrix $A$ from GOE ensemble; the normal complex Gaussian matrix $A$, which is known to have identical spectral pdf as complex non-normal Ginibre ensemble, (iii) 2 models based on non-normal Ginibre matrix, where the overlaps of eigenvectors play crucial role, exemplifying relation~(\ref{phi}).  \\
First (case (i)), we  assume that $A$ is normal and modeled by matrices from the (diagonally shifted) Gaussian Orthogonal Ensemble, $A=-\mu \idm+GOE$, characterized by the probability distribution)
\begin{equation}
P(A)dA\sim e^{-N\tr(A+\mu\idm )^2}dA.
\end{equation}
where we choose $\mu\geq 1$ to ensure stability of the corresponding dynamics. 
Assuming the simplest unstructured noise $B=\idm$ it is straightforward to see that the solution of \eqref{eq:CO} is given by $C_0=-\frac{1}{2}A^{-1}$. 
As the spectral density of $A$ in this case is a shifted semicircle $\rho_A(x)=\frac{2}{\pi}\sqrt{(1-\mu-x)(x+\mu+1)}$, a simple change of variables $y=-\frac{1}{2x}$  yields the density of eigenvalues for $C_0$ as:
\begin{equation}
\rho_{C_0}(y)=\frac{\sqrt{\mu^2-1}}{\pi y^3}\sqrt{\left(y-\frac{1}{2(\mu+1)}\right)\left(\frac{1}{2(\mu-1)}-y\right)}.
\end{equation}
The limit $\mu \to 1$ is finite, but the the spectrum of the covariance matrix becomes unbounded with the power-law tail:
\begin{equation}
\lim_{\mu\rightarrow 1}\rho_{C_0}(y)=\frac{1}{\pi y^3}\sqrt{y-\frac{1}{4}}
\label{eq:DensGOEcrit}
\end{equation}
This power-law behavior is related to the criticality of the system~\cite{Vinayak2014} and originates from the May's stability-complexity transition~\cite{May}.\\
Next (case (ii)),  we consider $A$ to be normal with complex eigenvalues, which we take from the class of shifted normal ensembles sharing the joint probability density of eigenvalues with the non-normal Ginibre matrices. Still keeping $B=\idm$, \eqref{eq:C0SpecDec} reduces now to
\begin{equation}
C_0=-\sum_{k=1}^{N}\ket{R_k}\frac{1}{2\re \lambda_k}\bra{R_k},
\end{equation}
Observe that the density  of complex eigenvalue for the chosen ensemble is uniform inside the unit circle, so that the associated density of real parts is given by the Wigner semicircle.  Indeed,  the imaginary part of the eigenvalue $\lambda=x+iy$ are limited, for any fixed real part $x$, by $\pm \sqrt{1-x^2}$, i.e. by the rim of the Ginibre circle. This in turn  implies that the spectrum of $C_0$ is exactly the same as in the formerly considered case of shifted GOE matrices.\\
Finally (case (iii)) we discuss two models with "tunable non-normality",  starting with matrices whose mean spectral density is given by the Ginibre law. In {\it Model 1} we begin with a Ginibre matrix $X$ with the  uniform limiting density of eigenvalues inside the unit disk. In the next step a Schur decomposition is performed $X=U(\Lambda+T)U^{\dagger}$. The upper-triangular part is rescaled as $T\rightarrow\eta T$ with $\eta\geq 0$ and the matrix $A$ is constructed as $A=U(\Lambda+\eta T)U^{\dagger}$. Finally, the matrix is shifted $A\rightarrow A-\mu\idm$ to ensure stability of the dynamical system. In {\it Model 2} the
matrix $A$ is composed by commencing with a Ginibre matrix $X$ whose elements within each row are constrained to sum up to 0 by imposing 
$X_{ij}\rightarrow X_{ij}-\frac{1}{N}\sum^N_{j=1}X_{ij}$. Then we define $A=-\mu\idm +X+\nu M$ such that $M=\frac{1}{\sqrt{N}} u m^T$, where $u^T=(1,1,...,1)$ and $m^T=(m_1,m_2,...,m_N)$ with elements of $m$ satisfying $\sum^N_jm_j=0$ (for convenience we assume also $\sum_jm_j^2=1$). Two parameters $\eta, \nu$ of the models control the degree of non-normality in $A$. At the same time, the limiting mean eigenvalue density is insensitive to the non-normality parameters.  \textit{Model 2} was originally proposed by Rajan-Abbott~\cite{RajanAbbott}, however only recently the role of non-orthogonal eigenvectors was discussed~\cite{Gudowska_neuro,Tarnowski_2020}. \\
To further analyze spectrum of the covariance matrix in {\it Model 1} at the edge of a stability/instability transition, we  investigated numerically tail distributions of the eigenvalues for $C_0$ for various parameters $\eta$.   The eigenvalue density of $C_0$ is well-described by a power-law, 
 demonstrating that the spectral density displays heavier tails in the critical regime with increase in non-normality of the matrix $A$. Note that for $\eta=0$ one gets $\alpha =2.5$ as expected from \eqref{eq:DensGOEcrit}, because it corresponds to the normal Ginibre matrices.\\
In {\it Model 2} the non-normality of matrix $A$ affects only the largest eigenvalue of $C_0$ the magnitude of which grows like $\nu^2$~(Supplemental Material~\cite{SUP}). \\
We move to the   Rajan-Abbott model of neuron networks~\cite{RajanAbbott}  
and numerically study  the FDT "violation parameter" $\tilde{\Delta}(\tau)$, whose analytic formula at large $N$ is not yet available. The results are presented  in Fig. 1. We remind  that for a symmetric matrix $A$ this contribution would simply vanish. Hence the pronounced transient behavior is due to growing non-normality parameter $\nu$.  
The observed non-orthogonality of eigenvectors is responsible for a rather complex transient behavior \cite{Tarnowski_2020,Grela_2017}  at intermediate temporal scales.  With non-normal  $A$ in that case played by the associated Jacobian matrix,  a perturbation to its entries affects eigenmodes and interferes positively to enhance susceptibility of the system to the perturbation, leading to increase in its capacity to exhibit a transient growth of deviation in linear regime \cite{Grela_2017}.  This enhancement of a system's reactivity has gained recently a broad attention and the effect has been discussed for large complex networks representing ecosystems \cite{Caravelli_2016}, spatio-temporal formation of Turing patterns \cite{Asllani_2018} and models of neural systems \cite{Gudowska_neuro,Marti,Duan_2022,Bialek}.
It turns out that the average entropy production rate in \textit{Model 2} can be calculated analytically in the limit of large system size (see Supplemental Material~\cite{SUP})  and reads
\begin{equation}
    \Phi = N  \left(1+\nu^2\right)\left(\mu + \sqrt{\mu^2-1}\right)^{-1}.
\end{equation}
It clearly shows that increasing non-normality in the system (controlled by $\nu$) enhances production of entropy.\\
One of the challenges of neuronal dynamics is the identification of the mechanism, that allows,  in non-equilibrium systems, the persistence   of long-lived, synchronized phenomena alike memory.  We suggest, that eigenvector correlations are natural candidates for such  behavior. To  strengthen our prediction, we have performed the simulation of the linearized dynamics  of neurons in Rajan-Abbott model, i.e. we have simulated 
\begin{equation}
    \frac{dx(t)}{dt}  = (-\mu{\bf 1} +X +\nu M) x
\end{equation}
for  $N$ neurons characterized by random spikes at $t=0$.  The results are presented in Fig. \ref{fig:RAsynchro}. 
Note, that by construction, all three simulations for different non-normality parameter $\nu$ share identical eigenvalues. Clearly, the difference can come only from eigenvectors. Non-zero   $\nu$  leads to approximate scaling of different neuronal trajectories,  and to the emergence of collective, oscillatory patterns. The strength of oscillations grows with $\nu$, alike does the entropy production. 

\begin{figure}[t!]
\begin{center}
\includegraphics[width=0.99\columnwidth]{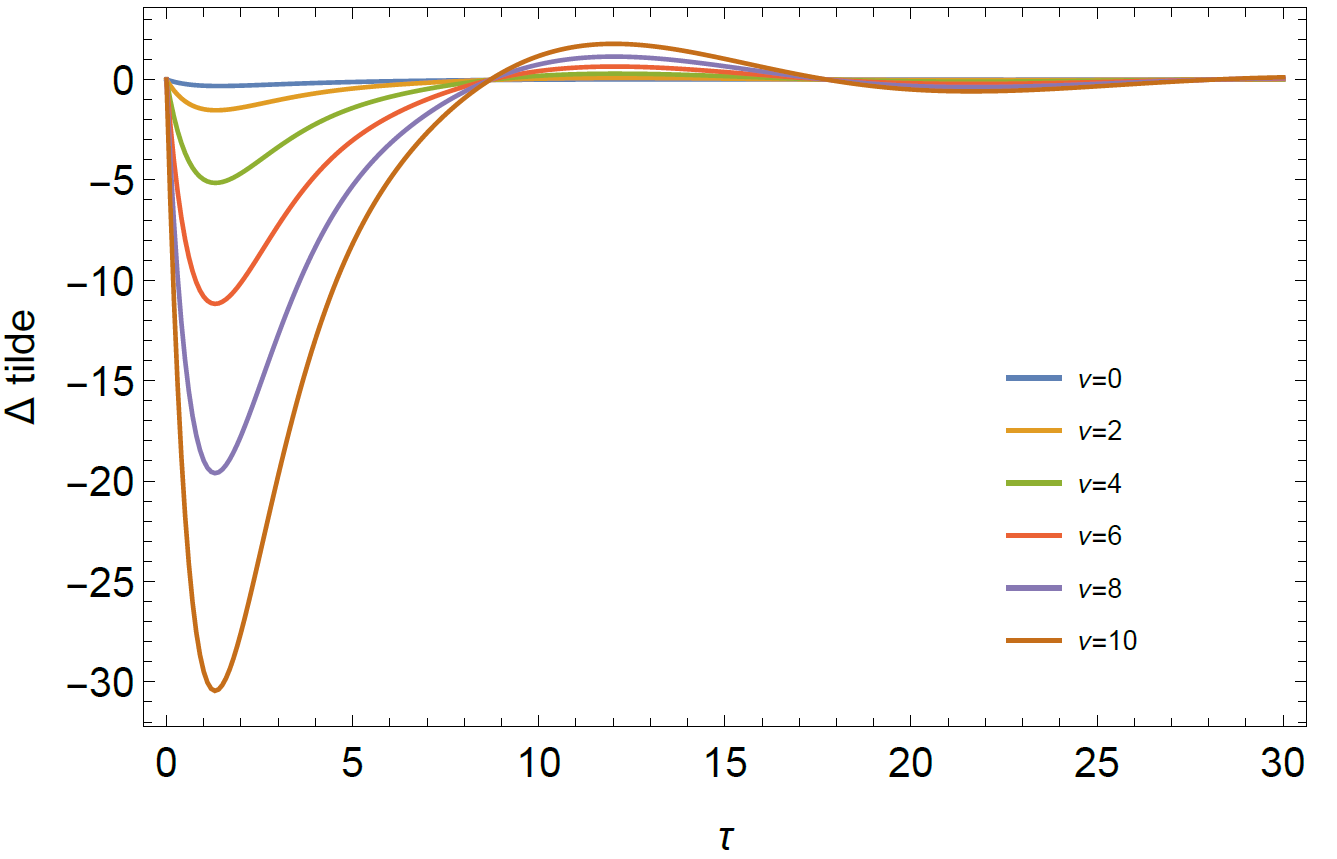}
 \end{center}
 \caption{ 
 Numerical results for $\tilde{\Delta}(\tau)$  for  various values of non-normality parameter $\nu$. 
 The transient behavior is dominant for intermediate time scales and is  enhanced by larger non-normalities. The "anomalous" contribution to FDT decays at large times.  }
 \label{fig:2}
\end{figure}

\begin{figure}[!t]
\begin{center}
\includegraphics[width=0.99\columnwidth]{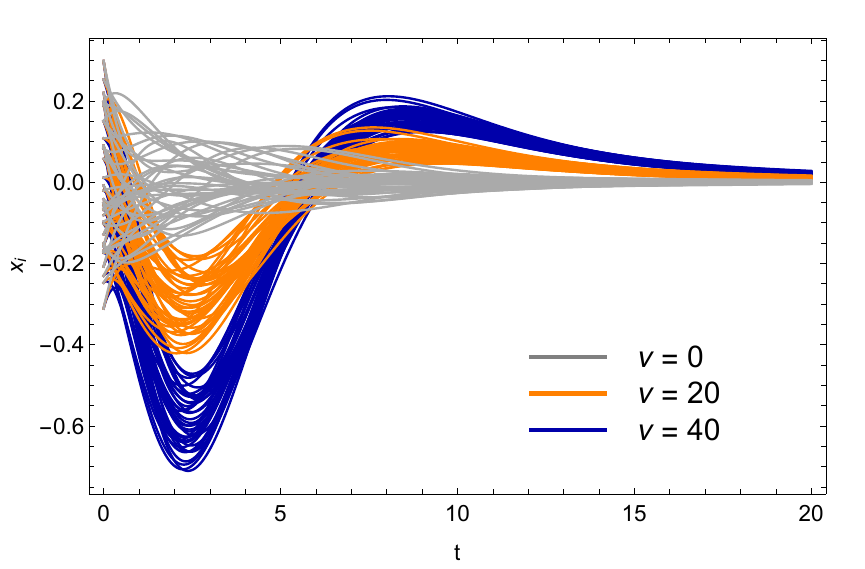}
 \end{center}
 \caption{
The activity as a function of time of each out of $N=40$ neurons in the linearized dynamics.  In all three cases ($\nu=0, 20, 40$ ) eigenvalues  are identical by construction. An onset of collective dynamics driven by the non-normality marker $\nu$  is clearly visible.   All simulations started from exactly the same initial condition randomly chosen from the N-dimensional hypersphere. For simplicity, the ratios of excitatory to  inhibitory neurons are identical. 
 }
     \label{fig:RAsynchro}
\end{figure}
\noindent In conclusion, in this work we have  pointed the crucial role of eigenvector nonorthogonality overlap matrix of the Chalker-Mehlig type for the production of entropy in non-equilibrium states  in the case of non-normal non-Hermitian operators governing a multivariable Langevin-type relaxation.  
 We suggested, that this particular mechanism  is responsible for the  emergence  of the transient dynamics in complex systems with  classical dynamics.  Taken into account the generality of our construction,  
 one is tempted to speculate  that importance of eigenvector overlaps should manifests itself also in features of quantum dynamics.  Indeed, series of works, including several very recent ones \cite{Savin,schomerus2000quantum,davy2019probing,fyodorov2022eigenfunction,cipolloni2023non,cipolloni2023entan,ghosh2023eigenvector} seem to corroborate this  suggestion.\\
  Finally, we tested these concepts on simple models  of neuronal systems, in which it is known that the increasing non-normality  of connectivity matrices leads directly to speeding up of system's dynamics~\cite{Marti}.
This observation is in agreement with  our earlier speculations~\cite{Gudowska_neuro}, that the possible synchronisation may come from the transient behavior, driven mainly by the coupled dynamics of complex eigenvalues and eigenvectors and agrees qualitativelly with the full numerical simulations of Rajan-Abbott models \cite{DelMolino}. We have also provided an explicit formula for the entropy production in Rajan-Abbott model, demonstrating explicit dependence on  non-normality parameter, and we have simulated the neuronal activity as a function of time and non-normality parameter. 
We hope that relatively simple form of the extended FDT and the recent outburst of  analytic results for  eigenvector correlations in random matrix theory  will facilitate  understanding of the main features of real neural networks.  This particular  fusion of ideas coming from statistical physics, random matrix theory and  neuroscience can be also applied to other domains of  complex systems, like so-called odd-diffusion, where collisions enhance the dynamics~\cite{Sharma2,Sharma1}. \\

\begin{acknowledgments}
The authors are indebted to Claude Godreche and Jean-Marc Luck for bringing their related work to the  authors' attention and to Abhinav Sharma for pointing the link  to the analysis of odd-diffusive systems. 
Y.V.F. acknowledges financial support from EPSRC Grant EP/V002473/1 ``Random Hessians and Jacobians: theory and applications''.
 This work was supported  by the TEAMNET POIR.04.04.00- 00-14DE/18-00 grant of the Foundation for Polish Science and by the Priority Research Area Digiworld under the program Excellence Initiative-Research University at the Jagiellonian University in Krakow.   
\end{acknowledgments}
\section*{Supplemental Material}

\renewcommand{\>}{\right>}

\subsection*{Relation between  Lyapunov (Sylvester) equation and  Einstein's form of FDT}  
To demonstrate the  physical content of Lyapunov (Sylvester) equation, let us recall a relevant example of a Brownian motion in a viscous fluid described by the Langevin equation $\dot{p}=-m\omega^2x -\gamma\frac{p}{m}+\xi(t)$ with $\dot{x} =\frac{p}{m}$. 
Identification of matrices reads: ${\bf{C}}_0={\rm diag}(<x^2>, <p^2>)={\rm diag} (k_BT/m\omega^2, mk_BT)$, where equilibrium equipartition is assumed. Then 

\begin{eqnarray}
{\bf A}=
\left( \begin{array}{cc} 0 & 1/m \\ -m\omega^2 & -\gamma/m
\end{array} \right) \,\,\,\,\,\,\,\,\,\,
{\bf B}=
\left( \begin{array}{cc} 0 & 0 \\  0 & 2\gamma k_B T
\end{array} \right) 
\end{eqnarray}
hence non-zero corner  of  matrix ${\bf B}$  provides  Einstein's form of the FDT relating the temperature, 
dissipation and diffusion.

\subsection*{Analysis of Lyapunov (Sylvester) equation in Fourier space}
\noindent In the the Fourier space $C(\omega)=\int_{-\infty}^{+\infty}e^{i\omega \tau}C(\tau)d\tau$ and the correlation function reads
\begin{eqnarray}
C(\omega)
=-(i\omega+A)^{-1}C_0+C_0(i\omega-A^T)^{-1}.
\end{eqnarray}
Using the explicit form of $C_0$ and the spectral decomposition of $A$ the above can be further rewritten as
\begin{equation}
C(\omega)=
-\sum_{kl}\frac{\ket{R_k}\bra{L_k}B\ket{L_l}\bra{R_l}}{(i\omega+\lambda_k)(i\omega-\bar{\lambda}_l)},
\end{equation}
which is equivalent to the product form
\begin{equation}
C(\omega)=(A+i\omega)^{-1}B(A^T-i\omega)^{-1}. \label{eq:CFourier}
\end{equation}
Following Eq.1 
the response function (or the general susceptibility of the system) with respect to the random field $\xi(t)$ is
$\chi(t)=e^{At}\theta(t)$ which in the Fourier space reads
$
\chi(\omega)=\int_0^{\infty}e^{At}e^{i\omega t}dt=-(i\omega+A)^{-1}.
$
Consequently, 
$
C(\omega)=\chi(\omega)B\chi^T(-\omega), 
$
and
\begin{equation}
i\omega\, C(\omega)=-B\chi^T(-\omega)-AC(\omega)=\chi(\omega)B+C(\omega)A^T.
\label{eq:C_w}
\end{equation}
The last identity yields 
\begin{equation}
2i\omega C(\omega)=\chi(\omega)B-B\chi^T(-\omega)+C(\omega)A^T-AC(\omega). \label{eq:FDTFourier}
\end{equation}
First, let us make a simplifying assumption of $A$ being normal. Then substituting in the above $B=2k_BT\idm$ brings about another form of the FDT relation: 
\begin{equation}
\chi(\omega)-\chi^T(-\omega)=\frac{i\omega}{k_BT} C(\omega)
\end{equation}
 Note that $\chi^{\dagger}(\omega)=\chi^T(-\omega)$, thus we have the anti-hermitian part ($\im (X)=\frac{X-X^{\dagger}}{2i}$) of $\chi(\omega)$ as a generalization of the imaginary part for the scalar case. Since $C(\omega)$ is a hermitian matrix (cf. (27)),
 we can write
 \begin{equation}
 \omega C(\omega)=\im (\chi(\omega) B+C(\omega) A^T).
 \end{equation}

\subsection*{Stochastic entropy and entropy production rate}
For a system described by the (multivariate) probability density $P({\bf{x}},t)$ we use a standard (Shannon's) definition of entropy
\begin{eqnarray}
  {\cal{S}}=-\int P({\bf{x}},t)\log P({\bf{x}},t)d{\bf{x}}  
\end{eqnarray}
and assume that $P({\bf{x}},t)$ satisfies the Smoluchowski-Fokker-Planck equation for the probability evolution along the stochastic paths:
\begin{eqnarray}
\partial_t P({\bf{x}},t)=\nabla\cdot[-{\bf{A}}({\bf{x}})P({\bf{x}},t)+\frac{{\bf{B}}}{2}\nabla P({\bf{x}},t)]
\label{FPE}
\end{eqnarray}
The change of entropy in time is then given by
\begin{eqnarray}
{\dot{\cal{S}}}=\int (\log P +1)\nabla\cdot{{\bf{J}}}dx=\Phi-\Pi_h\\
\end{eqnarray}
where
\begin{equation}
{{\bf {J}}}=-\frac{1}{2}{\bf{B}}\nabla P({\bf{x}},t)+{\bf{A}}({\bf{x}})P({\bf{x}},t)
\end{equation}
stands for  the probability flux,
\begin{equation}
 \Phi=-\int P({\bf{x}},t)^{-1}(\nabla P({\bf{x}},t)-2{\bf{B}}^{-1}{\bf{A}}({\bf{x}})P({\bf{x}},t))\cdot {\bf{J}}d{\bf{x}}   
\end{equation}
is identified as the entropy production rate and
\begin{equation}
\Pi_h=\int 2{\bf{B}}^{-1}{\bf{A}}({\bf{x}})\cdot{\bf{J}}d{\bf{x}}    
\end{equation}
represents rate of heat dissipation~  \cite{Zwanzig,Maes20,Fox1979,Godreche_2019}.
For a multidimensional Ornstein-Uhlenbeck process (cf. Eq.(1)) a drift (friction) matrix is ${\bf{A}}({\bf{x}})={\bf{A}\bf{x}}$ and $\bf{B}$ denotes $N\times N$ covariance matrix of the white Gaussian noise input
$<{\boldsymbol{\xi}}(t){\boldsymbol{\xi}}^T(t')>={\bf{B}}\delta(t-t')$. Here bold symbols refer to matrices and vectors.
The stationary (long time) solution to Eq.(\ref{FPE}) is
\begin{eqnarray}
    P({\bf{x}})=2\pi^{-N/2}(\text{det} {\bf{C}}_0)^{1/2}\exp(-\frac{1}{2}{\bf{x}}^T{\bf{C}}_0^{-1} {\bf{x}})
    \label{PSTAT}
\end{eqnarray}
from which one obtains $\nabla P({\bf{x}})={\bf{C}}_0^{-1}{\bf{x}}$.
 The entropy production rate at the stationary state takes the form of
\begin{eqnarray}
    \Phi=\frac{1}{2}\int(\nabla\log P({\bf{x}})-2{\bf{B}}^{-1}{\bf{Ax}})^T{\bf{B}}(\nabla\log P({\bf{x}})-\\ \nonumber
    -2{\bf{B}}^{-1} {\bf{Ax}})P({\bf{x}})d{\bf{x}}=\\ \nonumber
    =2\int \frac{{\bf{J}}^T({\bf{x}}){\bf{B}}^{-1}{\bf{J}}({\bf{x}})}{P({\bf{x}})}d{\bf{x}}
    \label{stat_EPR}
\end{eqnarray}
By making use of relation Eq.(\ref{PSTAT}) the stationary flux is ${\bf{J}}=\frac{1}{2}{\bf{B}}P({\bf{x}}){\bf{C}}_o^{-1}{\bf{x}}+{\bf Ax}P({\bf{x}})$. Inserting these expressions into Eq.(\ref{stat_EPR}) one gets
\begin{eqnarray}
    \Phi=<{\bf{x}}^T(2{\bf{B}}^{-1}{\bf{A}}-{\bf{C}}_0^{-1})^T\frac{{\bf{B}}}{2}(2{\bf{B}}^{-1}{\bf{A}}-{\bf{C}}_0^{-1}){\bf{x}}>
\end{eqnarray}
Now, observing that at the stationary state $<{\bf{x}}^T {\bf{Gx}}>=\rm{tr}({\bf{C}}_0{\bf{G}})$ and making use of 
 the FDT violation term ${\boldsymbol{\Delta}}(t)$ which allows us to express ${\bf{AC}}_0=({\bf{B}}+{\boldsymbol{\Delta}})/2$ 
 and ${\bf{C}}_0{\bf{A}}^T=({\bf{B}}-{\boldsymbol{\Delta}})/2$, 
we arrive at the final formula (cf. Eqs.(18,19) in the main text)
\begin{eqnarray}
\Phi= -\frac{1}{2}\rm{tr}({\bf{C}}_0^{-1}{\boldsymbol{\Delta}} {\bf{B}}^{-1}{\boldsymbol{\Delta}})=-{\rm tr }({\bf{B}}^{-1}{\bf{A}}{\boldsymbol{\Delta}}).   
\end{eqnarray}

 \subsection*{Entropy production rate in model 2}

Combining (17) with (11) and (7),
we can represent entropy production rate as
\begin{equation}
    \Phi = \tr \int_0^{\infty} A e^{As} e^{A^Ts} A^T ds - \tr \int_0^{\infty} A^2 e^{As} e^{A^Ts} A^T ds,
\end{equation}
where we already set $B=\idm$. The Dunford-Taylor integral~\cite{Dunford,Dunford2}, which is a matrix generalization of the Cauchy's integral formula, allows one to represent any analytic function of a matrix as
\begin{equation}
    f(X) = \frac{1}{2\pi i} \oint_{\mathcal{C}} \frac{f(z) dz}{z-X},
\end{equation}
where the integration contour $\mathcal{C}$ encircles all eigenvalues. Taking into account that in all models $A = -\mu \idm +X$, we rewrite the entropy production rate as
\begin{eqnarray}
    \Phi &=& \frac{1}{(2\pi i )^2} \int\limits_0^{\infty}ds \oint\limits_{\mathcal{C}}dz \oint\limits_{\mathcal{C}}
    dw\, (z-\mu)(w-z) \nonumber \\
    &\times&e^{s(z+w-2\mu)}  R_1(z,w), \nonumber
\end{eqnarray}
 where we introduced a traced product of resolvents
 \begin{eqnarray}
     R_1(z,w) =\tr \frac{1}{z-X} \frac{1}{w-X^T}.
 \end{eqnarray}
Since $X$ is random, we consider ensemble average of $R_1$ as the correlation function $W_1(z,w) = \< \frac{1}{N}R_1(z,w)\>$. The $\frac{1}{N}$ factor is introduced to obtain the finite expression in the large $N$ limit, which is a consequence of extensivity of entropy. This correlation function in large $N$ has been calculated in~\cite{TarnowskiBessel} and reads
\begin{eqnarray}
    W_1(z,w) = \frac{1}{zw-1}\left(1+\frac{\nu^2}{zw}\right).
\end{eqnarray}

We calculate the double contour integral first
\begin{eqnarray}
    &&\frac{1}{(2\pi i)^2}\oint\limits_{\mathcal{C}}dz \oint\limits_{\mathcal{C}}
    dw\, (z-\mu)(w-z) e^{s(z+w-2\mu)} \nonumber \\
    &\times&\frac{1}{zw-1}\left(1+\frac{\nu^2}{zw}\right).
\end{eqnarray}
The integral over $w$ can be evaluated by the method of residues. There are two residues $w_1=1/z$ and $w_2=0$. The second residue, however, yields an integral that is analytic in $z$, hence integral over $z$ yields 0. The residue at $w_1$ leads to
\begin{eqnarray}
    \frac{1+\nu^2}{2\pi i} \oint\limits_{\mathcal{C}}dz\, \frac{z-\mu}{z}\left(\frac{1}{z}-z\right)e^{s\left(z+\frac{1}{z}-2\mu\right)}.
\end{eqnarray}
If we denote by $\tilde{\Phi} = \frac{1}{N}\<\Phi\>$ the average entropy production rate per component of the system, it is therefore given by
\begin{equation}
    \tilde{\Phi} = (1+\nu^2)\int_0^{\infty}e^{-2\mu s} \left(I_0(2s) - I_2(2s)\right) ds, \label{eq:EntrProd}
\end{equation}
where we used the integral representation of modified Bessel function of the first kind
\begin{eqnarray}
    I_a(s) = \frac{1}{2\pi i} \oint_{\Gamma} e^{\frac{s}{2}\left(z+\frac{1}{z}\right)} \frac{dz}{z^{a+1}},
\end{eqnarray}
where the integration contour $\Gamma$ encircles $z=0$. The integrals in \eqref{eq:EntrProd} are calculated using a well known formula
\begin{equation}
    \int_0^{\infty} e^{-\mu s} I_a(s)ds = \frac{1}{\sqrt{\mu^2-1}\left(\mu+\sqrt{\mu^2-1}\right)^a},
\end{equation}
which finally results in
\begin{equation}
    \tilde{\Phi} = \left(1+\nu^2\right)\left(\mu + \sqrt{\mu^2-1}\right)^{-1}.
\end{equation}
Note that putting $\nu^2=0$ yields  the entropy production rate  in the Ginibre ensemble, since for this ensemble $W_1(z,w)=\frac{1}{zw-1}$, as a special case for so-called R-diagonal operators~\cite{NowakTarnowski}.

Interestingly, only the largest eigenvalue of $C_0$ in model~2 is affected by non-normality of matrix $A$. 
This is visible on the Figure~3 here, where 4 largest eigenvalues are shown  on log-log plot as a function of $\nu$.  

 \begin{figure}[!h]
\begin{center}
\includegraphics[width=0.99\columnwidth]{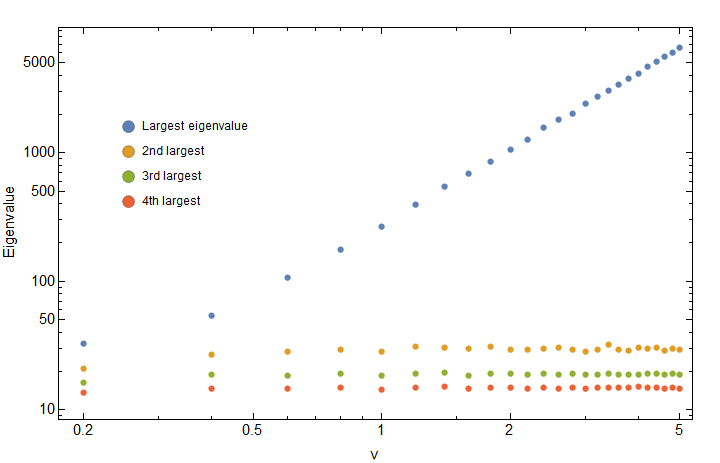}
 \end{center}
 \caption{ Four largest eigenvalues of $C_0$ in {\it Model 2} as a function of the non-normality parameter $\nu$. Only the largest one is affected, while the density of other eigenvalues remains almost unchanged.}
     \label{fig:zipf}
\end{figure}

 \section*{References}

\bibliography{bibliography}
\end{document}